\documentstyle[12pt]{article}

\begin{document}

\begin{centering}

\begin{flushright}
gr-qc/0207049 $~~~~~~~~~$\\
preprint version\footnote{The published version was edited
and shortened. The title was changed to ``Special Treatment",
with ``Relativity" as subject heading,
and a figure (with elementary description of cosmic-ray observations)
was added.} of $~~~~~$\\
Nature 418 (2002) 34-35
\end{flushright}

\bigskip
{\Large \bf Doubly Special Relativity}\\
\bigskip
\bigskip
\bigskip
{\bf Giovanni AMELINO-CAMELIA}\\
\bigskip
Dipartimento di Fisica, Universit\`{a} ``La Sapienza", P.le Moro 2,
I-00185 Roma, Italy
\end{centering}
\vspace{1cm}

\vspace{0.5cm}
\begin{center}
{\bf ABSTRACT}
\end{center}

\begin{center}
\begin{centering}
\begin{minipage}{4.5in}\footnotesize\baselineskip=10pt
I give a short non-technical review of the results
obtained in recent work on ``Doubly Special Relativity",
the relativistic
theories in which the rotation/boost transformations between
inertial observers are characterized by two
observer-independent scales (the familiar velocity scale, $c$,
and a new observer-independent length/momentum scale, naturally
identified with the Planck length/momentum).
I emphasize the aspects relevant for the search of
a solution to the cosmic-ray paradox.
\end{minipage}
\end{centering}
\end{center}

\vspace{0.5cm}

\baselineskip = 12pt

Galilei/Newton Relativity was abandoned because of the conflicting
results of the Michelson-Morley experiments
and because of its incompatibility with the mathematical structure
of the successful Maxwell theory of electromagnetism.
After a century of successes,
Einstein's Special Relativity, the theory that replaced
Galilei/Newton Relativity, could now be questioned for similar reasons.
A key Special-Relativity prediction appears to be violated
by certain observations of ultra-high-energy
cosmic rays~\cite{gzkdata}, and some Quantum-Gravity arguments
appear to encourage a modification of Special Relativity.
I argued in Ref.~\cite{dsr1} that this situation can provide motivation
for considering a change in the Relativity postulates somewhat
analogous to the
one needed for the transition from Galilei/Newton Relativity
to Einstein's Relativity: the introduction of a new absolute
observer-independent scale.
I analyzed in detail a first illustrative example of relativity
postulates with an observer-independent length(momentum) scale,
in addition to the familiar observer-independent velocity scale $c$,
and I showed that there is no {\it in principle}
obstruction for the construction of such new
relativistic theories.
That illustrative example of new
relativistic theory also predicts some
new effects of a type that could explain the observations
of ultra-high-energy cosmic rays, but the predicted magnitude
of these effects turns out to be too weak for a description
of the data.
For nearly a year, follow-up studies~\cite{follow}
focused on the example of new relativistic theory
which I had used to illustrate the idea.
In a very recent paper~\cite{leedsr}, Maguejio and Smolin
constructed a second example of relativistic theory of the
type proposed in Ref.~\cite{dsr1},
and this has generated increased
interest~\cite{dariofrancesco,jurekDSRnew,lukiedsr,judesVisser,AKspinors,dsrLODZ,granik}
in research on relativity theories with two observer-independent
scales. These theories are being
called\footnote{{\bf Note added}: Of course,
the name ``Doubly Special Relativity"
was not the only possible choice.
In particular, in a very recent paper~\cite{AKspinors},
reporting interesting results on fermions and bosons
in DSR, it was argued that the
less-compact name ``Special Relativity with two invariant scales"
could be preferable.
Since the name ``Doubly Special Relativity" has already
been adopted by a few research
groups~\cite{dsr1,jurekDSRnew,lukiedsr,judesVisser,dsrLODZ,granik}
it is probably best at this point to maintain it, avoiding
the risk of a two-name confusion in the literature.
Moreover, the name ``Special Relativity" has acquired a status
in the literature which goes beyond the intended meaning:
it is the specific theory proposed by Einstein, which in particular has one
observer-independent scale. The
name ``Special Relativity with two invariant scales" may be confusing
since Special Relativity is identified with the known theory
with one observer-independent scale.
It would also be awkward to adopt the corresponding
name ``Special Relativity without invariant scales"
for the Galilei/Newton theory.
It is too early to think of a name
to be adopted in the event of the proposal of a deformation of
General Relativity which would be consistent with the DSR deformation
of Special Relativity, but in perspective I notice that
(assuming the description of $\hbar$ reserves no surprises)
such a deformation of GR should not require the introduction
of an extra scale ($E_p$ is already there as a coupling constant):
it might require attributing~\cite{dsr1,dsrPOLON} to $E_p$
the double role of both kinematical scale and coupling constant.}
``Doubly Special Relativity" (DSR) theories.
Some of the key open issues that are being studies concern
the search of other examples of DSR theories,
the study of the multi-particle sector of given DSR theories,
and attempts of finding a full solution of the cosmic-ray
paradox within DSR.

An aspect of cosmic-ray observations that is directly connected with
relativity is the
Greisen-Zatsepin-Kuzmin (GZK) limit. Ultra-high-energy
cosmic rays are particles,
most likely protons, produced by distant active galaxies, which we detect
through the particle-physics processes they ignite in the atmosphere. The
GZK limit is related with the threshold energy,
$E_{GZK}$, required for such cosmic rays to interact with
the ``Cosmic Microwave Background Radiation" (CMBR):
because of these interactions with the CMBR
it should not be possible for cosmic rays with
energies above $E_{GZK}$ to reach us. The value of the threshold energy
$E_{GZK}$ is a purely kinematical prediction. In a given relativity
theory it is obtained by combining the laws of energy-momentum conservation
and the dispersion relation (the relation between the energy
and the momentum of a particle). Within Special Relativity
one finds $E_{GZK} \simeq 5 {\cdot} 10^{19} eV$, but several
cosmic-ray events~\cite{gzkdata} are in disagreement
with this prediction. As with all emerging experimental paradoxes
it is of course possible that the
cosmic-ray paradox is the result of an
incorrect analysis of the experiment, for example it is legitimate to
speculate that the identification of these ultra-high-energy cosmic rays as
protons produced by distant active galaxies might eventually turn out to be
incorrect. But, in spite of its preliminary status,
this cosmic-ray paradox provides encouragement for
the study of new relativity postulates.

Besides this motivation coming for the experimental side,
the idea of revising the relativity postulates also finds encouragement
because of the role that the ``Planck scale" $E_p \simeq 10^{28} eV$
plays in certain Quantum-Gravity scenarios.
Various arguments lead to the
expectation that for particles with energies close to $E_p$
it would be necessary to describe spacetime in terms of one form or
another of new spacetime quanta, while for our readily
available particles with energies much smaller than $E_p$
the familiar classical-spacetime picture would remain valid.
It would be appealing to introduce such a transition scale, a scale
at which our description of physical phenomena changes
significantly, as an observer-independent kinematical scale.

Just like in going from Galilei/Newton Relativity to Einstein's
Special Relativity
a key role is played by the law of composition of velocities,
which for Einstein must reflect the special status of the speed-of-light
constant $c$, in going then from Special Relativity to DSR
a key role is played by the law of composition of energy-momentum,
which, in the new framework, must attribute a special status to
the Planck scale $E_p$.
This modification of the laws of energy-momentum conservation
also affects the evaluation of the mentioned
threshold energy $E_{GZK}$ that is relevant for the cosmic-ray paradox.
Indeed the DSR theory
which I considered as illustrative example in Ref.~\cite{dsr1}
predicts, besides the emergence of a maximum momentum of order $E_p$
for fundamental particles,
a deformation of the laws of energy-momentum conservation.
In turn the deformed laws of energy-momentum conservation
of the DSR theory of Ref.~\cite{dsr1} leads to
a value of $E_{GZK}$ which is different from the one predicted
by ordinary Special Relativity, but the difference
is not as large as required for a full explanation
of the cosmic-ray paradox.
The DSR theory more recently proposed~\cite{leedsr}
by Maguejio and Smolin predicts that $E_p$ sets the maximum
value of both energy and momentum for fundamental particles.
The corresponding deformation~\cite{dariofrancesco} of the
laws of energy-momentum conservation
leads to yet another alternative prediction for $E_{GZK}$,
but that too is not sufficient to solve the cosmic-ray paradox.

A lot has been understood of these DSR theories, but several
issues must still be investigated.
At present, since the status of the paradox is still preliminary,
the search of a solution to the cosmic-ray paradox
is not to be seen as a consistency condition for the DSR research
programme, but the emergence of a compelling phenomenological result
would legitimate interest in
the rather speculative DSR idea.
The recent study~\cite{leedsr} by Magueijo and Smolin,
by showing that there exist more than one example of DSR
theories, provides indirect encouragement for the search of
a DSR theory that would solve the cosmic-ray paradox.
It appears~\cite{dsr1,dariofrancesco,dsrPOLON}
that a DSR theory capable of solving the cosmic-ray paradox
should rely on the introduction of some structure not yet present
in the DSR theories discussed in Refs.~\cite{dsr1} and~\cite{leedsr}.
This new structure should enter the law of composition of energy-momentum,
which, through energy-momentum conservation, affects the
evaluation of $E_{GZK}$.

Interestingly, the law of composition of energy-momentum
also plays a role in another open problem for DSR theories:
while Planck-scale deformed dispersion relations, generically predicted
in DSR~\cite{dsr1,leedsr}, are fully consistent with our
observations of fundamental/microscopic particles,
they are clearly in conflict with observations of macroscopic bodies.
Therefore the rule that defines the total momentum of a macroscopic
body in terms of the momenta of the composing microscopic particles
must allow the emergence of different relativistic
properties for macroscopic and microscopic
entities\footnote{{\bf Note added}: In particular,
it appears plausible that we might have
to assign to the ``observer" (a key entity in physics,
but a rather abstract aspect of ordinary Special Relativity)
the properties of a macroscopic body. Therefore,
as long as we lack a satisfactory understanding of
macroscopic bodies in DSR, certain considerations
involving ``the observer", such as the considerations reported
in Ref.~\cite{dsrLODZ}, should be postponed.
The analysis attempted in Ref.~\cite{dsrLODZ} is interesting
but premature in DSR: it should be seen as raising important
issues for the DSR research programme
(particularly for the DSR scheme of Ref.~\cite{leedsr}),
rather than as a definitive study of those issues.}.
Such a difference between macroscopic and microscopic entities
could not be accommodated in ordinary Special Relativity,
but, remarkably, through the presence of an observer-independent
energy/momentum scale, DSR naturally provides room~\cite{dsr1,dsrPOLON}
for a macro/micro separation. Unfortunately, it is still
unclear in which specific way this macro/micro separation
would be best implemented.
In particular, it is unclear whether or not we should assume
that the total energy-momentum of a multi-particle system
is constructed in terms of the
energy-momentum of each composing particle according to laws
that are themself observer-independent laws.
In addition to the already difficult task of describing a system
of free particles both in terms of the momenta
of each particle and in terms of the total momentum of the system,
even more conceptually challenging for DSR theories is the
description of the relativistic properties of bound states
(macroscopic particles which are in a sense composed of
several fundamental particles, but the composing particles are
not free).
Besides the indirect connection with the cosmic-ray paradox through
the laws of composition of energy-momentum, this issue of the
macro/micro separation could turn out to be even directly relevant
for the analysis of the cosmic-ray paradox if it is established that
(some of) the relevant particles (protons, pions, photons)
must be described as composite objects in DSR.

The role of certain $\kappa$-Poincar\'{e} Hopf algebras~\cite{kappa}
in DSR theories is also a lively subject
of investigation~\cite{dsr1,jurekDSRnew,lukiedsr,dsrPOLON}.
In the one-particle sector of the DSR theories so far considered
the Lorentz sector of certain $\kappa$-Poincar\'{e} Hopf algebras
plays a role that is analogous\footnote{Since we call ``Lorentz
transformations" the one-particle-sector laws of transformation
between observers of Einstein's Special Relativity, it might
be appropriate to call ``$\kappa$-Lorentz transformations"
the one-particle-sector laws of transformation
between observers in those DSR theories in which
the Lorentz sector of $\kappa$-Poincar\'{e} Hopf algebras
has the corresponding role.}
to the role of the Lorentz algebra in Einstein's Special Relativity:
the operators of the (Hopf) algebra are used to describe infinitesimal
rotation/boost transformations between inertial observers.
But for what concerns the delicate issues just mentioned above, the
ones that involve the two-particle and multi-particle
sectors, the role of $\kappa$-Poincar\'{e} Hopf algebras
remains unclear~\cite{dsr1,dsrPOLON}.
It would also be interesting to establish whether it is possible to
construct DSR theories without
any use of $\kappa$-Poincar\'{e} Hopf algebras, not even for the
infinitesimal transformations of the one-particle sector.
In fact, the DSR proposal~\cite{dsr1} just requires that
the laws of rotation/boost transformation between inertial
observers have a lenght(momentum) and a velocity observer-independent
scales, and it might be possible to realize this proposal in
a variety of mathematical frameworks.

The debate on these open issues is likely to keep busy the interested
scientists for the next few years.
Experimental help could come soon from improved data on
the mentioned ultra-high-energy
cosmic rays; moreover, in 2006, when related studies~\cite{grbgac}
will be performed on the GLAST space telescope,
we should have conclusive
information on another key prediction of some DSR theories:
a possible wavelength dependence~\cite{dsr1,dsrPOLON}
of the speed of photons.

\vfil
\eject

\baselineskip 12pt plus .5pt minus .5pt

\baselineskip = 12pt

\vfil

\end{document}